\RequirePackage{fix-cm}
\documentclass[smallextended]{svjour3}       % onecolumn (second format)
\smartqed  % flush right qed marks, e.g. at end of proof
\usepackage{graphicx}
\usepackage{lineno,hyperref}
\usepackage{float}
\usepackage{multirow}
\usepackage{xcolor}
\usepackage{subfigure}
\begin{document}

\title{Neutrality May Matter: Sentiment Analysis in Reviews of Airbnb, Booking, and Couchsurfing in Brazil and USA\thanks{Grants that supported this study: CAPES - Finance Code 001, CNPq Grant \#403260/2016-7, FAPESP GoodWEB project Grant \#2018/23011-1.}
}
\titlerunning{Neutrality May Matter}



\author{Gustavo Santos \and
       Vinicius F. S. Mota \and
       Fabr\'icio Benevenuto \and
       Thiago H. Silva
}


\institute{Gustavo Santos \at
              Universidade Tecnológica Federal do Paraná - Curitiba, Brazil \\
              %\email{gustavosanto@alunos.utfpr.edu.br}           %  \\
           \and
           Vinicius Mota \at
              Federal University of Espirito Santo - Vitoria, Brazil\\
              %\email{vinicius.mota@inf.ufes.br}
            \and 
            Fabr\'icio Benevenuto \at
            Universidade Federal de Minas Gerais - Belo Horizonte, Brazil\\
            \email{fabricio@dcc.ufmg.br}\\
            \and
            Thiago H Silva \at
            \emph{Current address:} University of Toronto -  Canada\\
            \email{th.silva@utoronto.ca}
}

\date{Received: date / Accepted: date}

\maketitle

\begin{abstract}
Information and communications technologies have enabled the rise of the phenomenon named \textit{sharing economy}, which represents activities between people, coordinated by online platforms, to obtain, provide, or share access to goods and services. In hosting services of the sharing economy, it is common to have a personal contact between the host and guest, and this may affect users' decision to do negative reviews, as negative reviews can damage the offered services. To evaluate this issue, we collected reviews from two sharing economy platforms, Airbnb and Couchsurfing, and from one platform that works mostly with hotels (\textit{traditional economy}), Booking.com, for some cities in Brazil and the USA. Trough a sentiment analysis, we found that reviews in the sharing economy tend to be considerably more positive than those in the traditional economy.  This can represent a problem in those systems, as an experiment with volunteers performed in this study suggests. In addition, we discuss how to exploit the results obtained to help improve users' decision making.
\end{abstract}


%\linenumbers

\section{Introduction}

Recent advances in information and communications technologies have favored the rise of the so-called \textit{sharing economy}. The sharing economy represents collaborative (person-to-person) activities to obtain, provide, or share access to goods and services, coordinated by online services based on a community of users \cite{hamari2016sharing}. Platforms of this new type of economy have conquered a significant market share in several segments such as transportation with, for example, Uber \footnote{http://www.uber.com.} and Cabify \footnote{https: // www.cabify.com.}, and hosting with, for instance, Airbnb \footnote{https://www.airbnb.com.}, and Couchsurfing \footnote{https://www.couchsurfing.com.} \cite{allen2015sharing}.

In the hosting market, Airbnb is a service that connects people who have a space to share with people who are looking for a place to stay, and the value of hosting is stipulated by the host \cite{zervas2014rise}. Couchsurfing allows people to share their areas in a similar way to Airbnb; however, the hosts do not charge for the service provided \cite{Tan:2010}. These services usually compete with hotels, inns, and even real estate companies, which have always been the dominant ones of this market, being, then, representatives of the traditional or formal economy.

Users on sharing economy platforms are typically invited to express their opinions about the service being used. These opinions can be captured in a variety of ways, among them, ratings and reviews (i.e., evaluative comments). In fact, these opinions are critical to many platforms in this segment. Companies like Uber require drivers to keep a certain feedback rating \cite{rogers2015social}. In the hosting context, negative opinions about hosts can impact the decision of future locations \cite{fradkin2015bias}. 

It is important to highlight that in hosting services of the sharing economy, there is personal contact between the host and the guest. At Airbnb and Couchsurfing, it is not uncommon for a guest to share accommodation with the host. This meeting may favor the creation of a relationship between who offers and hires the service, which does not tend to happen in hosting services of the traditional economy. Due to these personal contact,  guests may be in an uncomfortable position to perform a negative evaluation of services offered in the sharing economy, which may undermine an adequate assessment of the service consumed \cite{sparks,Gidumal2019}. Excessive positive reviews could also be favored by the fact that on Airbnb and Couchsurfing the rating systems are bidirectional, i.e., hosts rate guests, and guests rate the hosts.  In contrast, in the traditional economy, the rating is typically performed only by the guest; this is the case for Booking.com. This type of structural difference in the reputation mechanism has been hypothesized to explain more positive evaluations \cite{dellarocas2003digitization,dellarocas2010online}. These are some of the possible reasons to induce positive evaluations on the hosting services of the sharing economy, but, indeed, others could also be playing a role.

Based on those points, an important issue to be investigated is: do hosting reviews tend to be less negative in the sharing economy? Understanding this issue is crucial, as users' opinions are typically taken into account in the decision-making \cite{10.1007/978-3-211-77280-5_4,lee2008effect,pang2008opinion,dellarocas2010online}. The correct understanding of what reviews really mean can help in the construction of recommendation systems and ranking of services offered, which can help users to make better choices.

Our contributions to evaluate this issue can be summarized as follows:

\begin{itemize}
    \item We collected reviews from two sharing economy platforms, Airbnb and Couchsurfing, and a representative of the traditional economy, Booking\footnote{https://www.booking.com.}, a popular Web service to find hotel accommodations. We consider accommodations offered in three Brazilian cities and three cities in the United States;
    
    \item In possession of these reviews, we perform sentiment analysis on the shared texts. We find that reviews in the sharing economy tend to be more positive than those in the traditional economy. Besides, we present some key features of these comments, which reinforce the insights observed.
    
    \item  We performed a study with volunteers to evaluate how the observed phenomenon affects the user decision-making process. Our results suggest that the classification of establishments at Airbnb made by users could be affected due to the lack of negative evaluations. Our study still discusses how to explore the results obtained to assist in the decision-making of choosing accommodation in the sharing economy.
\end{itemize}

The remainder of the paper is organized as follows. Section \ref{secTrabRel} presents  related works. Section \ref{secDatasetsPlatform} describes the platforms evaluated, and the databases studied. Section \ref{secSent} discusses the concept of sentiment analysis and how we accomplish this task. Section \ref{secResultados} discusses the results obtained regarding the sentiment polarity in the analyzed systems, as well as some of the main characteristics related to the content and other factors. Section \ref{secAvaliacaoUsuarios} presents the results obtained with an experiment performed with volunteers to validate our results. Section \ref{secImplicacoes} proposes a  new way of evaluating hosting on the sharing economy. Finally, Section \ref{secConclusao} concludes the study and presents future work.
%~~~~~~~~~~~~~~~~~~~~~~~~~~~~~~~~~~~~~~~~~~~~~~~~~~~~~~~~~~~~~~~~~~~~~~~~~~~~~~~~~~% 

\section{Related Work} \label{secTrabRel} 

Recent research efforts have attempted to characterize and understand to what extent human language is biased towards positivity or negativity \cite{dodds2015human}. While there is controversy on this topic \cite{garcia2015language,dodds2015reply} there are spaces in the Web full of negativity. A key example is the comment section of online newspapers, which are likely to attract negative comments independently of the content of the news~\cite{reis2015@icwsm}. 

An online space in which opinions have become critical for the development of valuable tools is product reviews. Indeed, Pang et al. \cite{pang2008opinion} observe that ratings and opinions of other users in products are increasingly important in consumer choice. Not surprisingly, many efforts were dedicated to proposing a review summary and comparisons out of a large dataset of reviews \cite{hu2004mining,liu2005opinion,liu2012sentiment}. 

More recently, a new wave of efforts has attempted to extract opinions out of social media data, using mostly Twitter as a data source. Applications vary widely, from inferring political polls \cite{o2010tweets}, and inferring stock marketing fluctuations based on Twitter reactions \cite{bollen2011twitter}, to the
extraction of urban perceptions \cite{santosWI2018}. In common, most of these efforts rely on methodologies that exploit one out of the many existing methods for sentiment analysis to infer opinions \cite{Ribeiro2016,araujo2013measuring,araujo2016@icwsm}. 

Despite the undeniable advance that existing efforts have made on this space, all those efforts are devoted to exploring a large set of opinions that users express freely towards products, politicians, or companies in the case of stock marketing.  Our work explores opinions in a novel environment, in which the target of the review is a service or a product that involves some level of personal relationship, such as renting someone’s place for a short period.  

There are only a few efforts in the literature related to these kinds of environments. Particularly, Fradkin et al. \cite{fradkin2015bias} studied reviews and ratings on Airbnb websites and found that, on average, 72\% of users write at least one review of the place they stayed. The authors also found that 94\% of star ratings ranged from 4 to 5 (the scale goes from 1 to 5), which is important to show that, in fact, considering only the stars at one location for decision making may not be the best strategy. This is consistent with the results of Zervas et al. \cite{zervas2015first}, reporting that the average rating on Airbnb to be $4.7$. To get a better understanding of possible reasons for this phenomenon, Bulchand-Gidumal and Melián-González \cite{Gidumal2019} performed a study using surveys and interviews to investigate if guests faithfully convey their experiences on Airbnb. They found that a significant part of guests did not tell the whole truth when they evaluated, or avoid review when the experience was not positive. Some of the most important reasons for these behaviors include avoiding harming the host. Researchers have also studied a similar problem in the context of restaurants \cite{GANU20131}.

Fu et al. \cite{7023329} analyzed reviews on real estate, aiming to develop methods to classify properties according to market value. This could provide decision support for real estate buyers and, thus, can play a strategic role in the real estate market. Oliveira and Bermejo \cite{oliveira2017midias} applied sentiment analysis in social media data in the context of social and political management. This is important because traditional media landscape has changed considerably; in the past, traditional media was predominant (e.g., newspapers, magazines, and TV), and, recently, they are being complemented or replaced by online interactive social communication \cite{sobkowicz2012opinion}. Leung \cite{leung2009sentiment} uses sentiment analysis in product reviews. Leung shows that through sentiment analysis, it is possible to use the measured sentiment to support a product or give directions for improvements.

More related to the context of sentiment analysis in online hosting services, Duan et al. \cite{6480220} decompose user reviews into five dimensions envisioning to measure the service quality of a hotel. The authors used those dimensions in econometric models to study their effect on influencing users' behavior regarding the content generation and evaluation. Bridges and Vásquez \cite{Bridges2018} inspected the language used in 400 Airbnb reviews finding that 93\% of them were categorically positive.  Tian et al. \cite{tian2016mining} analyzed reviews of some three to five star hotels in China. They found that the star rating (given in the review) correlates well with the sentiment scores for both the title and the full content of the review. Mankad et al.  \cite{Mankad2016} studied reviews shared in hotels located in Russia and found, among other things, that negative reviews (negative sentiment) have a more significant downward impact than positive reviews.

To the best of our knowledge, our work differs from all previous related work because our focus is to study whether reviews on online hosting platforms on the sharing economy tend to be less negative than their competitors in the traditional economy. Also, we examined some of the key features of the comments we analyzed, performed user experiments to reinforce our findings, and discussed the implications for designing new features for sharing economy platforms.

%~~~~~~~~~~~~~~~~~~~~~~~~~~~~~~~~~~~~~~~~~~~~~~~~~~~~~~~~~~~~~~~~~~~~~~~~~~~~~~~~~~% 

\section{Data and Platforms Studied}\label{secDatasetsPlatform}

This section is divided into two parts: Section \ref{secPlataformas} presents the platforms studied, whereas Section \ref{secBase} describes the data collected from them.

\subsection{Platforms Considered}\label{secPlataformas}

We considered in this study three online hosting platforms. Booking is a traditional representative where hotels are the most common accommodations. In this platform, monetary payment is always demanded, and employees typically do the negotiation of the services, that is, without having personal contact with the owner of the business. Besides, we consider a platform that does not charge for the accommodations: CouchSurfing. In this platform, the negotiation is usually done by the owner of the lodging, and the personal contact between guest and host is high. Finally, we consider an intermediary representative: Airbnb. On Airbnb, the payment for the lodging is required, but the personal contact between host and guest tends to be high, often the accommodation is shared with the host.

\subsubsection{Airbnb}

Airbnb, founded in 2008, is a platform for private accommodation rentals around the world. Present in 190 countries and more than 34 thousand cities, it currently has more than 2 million accommodations \cite{airbnbAbout}. Airbnb economically empowers millions of people worldwide to open and capitalize on their spaces, becoming entrepreneurs of the hosting area. This platform has helped many travelers to save money, as it might be cheaper than hotels. Also, differentiated and more personal experiences are other factors that might help explain the success of this platform.
  
\subsubsection{CouchSurfing}

CouchSurfing (CS), created in 2004, is an online hosting platform, similar to Airbnb. It has served more than 11 million travelers in over 150,000 cities around the world \cite{couchsurfingAbout}. The differential of this service is that the accommodations are free and, typically, the personal contact between the host and the guest is higher than in Airbnb and Booking. While on Airbnb, users can share accommodation with hosts, in CS, this happens at practically every time.

Numerous features are available such as detailed personal profiles, an identity verification system, a personal certification system, as well as a personal referral system to increase security and trust among members. Perhaps because it does not involve monetary values, the user profile has a great value in this platform. Candidates for a poorly rated or dubious profile may find it more difficult to find accommodation than users with a flawless profile.

\subsubsection{Booking}

Booking, founded in 1996, is now one of the largest online hosting companies in the world. It has more than 15,000 employees in 198 offices in 70 countries around the world. This hosting service aims to connect travelers to various accommodation options, from small family-run inns to large 5-star hotels. Booking also offers options for private homes, similar to Airbnb. Despite the growth of this type of alternative, the majority of available accommodations, about 68.5\%, are hotel rooms. Of the remainder, 8.5\% are temporary rental options (``Vacation Rental Rooms", where several Airbnb style options are available), and 23\% are unique hosting options (``Unique Categories of Places to Stay"), where you can find unusual accommodations \cite{bookingStats}.

This company represents one of the most popular hosting sites in the traditional economy. Every day, more than 1,550,000 accommodations booked through its platform \cite{bookingAbout}.

\subsection{Dataset Describtion}\label{secBase}

We collected reviews of the systems considered (Airbnb, Couchsurfing, and Booking) for the cities of Curitiba, Rio de Janeiro, and S\~ao Paulo, Brazil, and the cities of Boston, Las Vegas and New York, in the United States. We chose these cities because they are popular with tourists and offer different types of attractions, possibly attracting different tourist profiles. For example, Curitiba is the third most visited city by foreigners for business tourism in Brazil \cite{curitibaTurismo}.

For each service, we first look for accommodations in the selected cities. We then collect all reviews made by users on all available hosting options. With this, our dataset is composed of an establishment identifier, a user identifier that made a review, as well as the review text itself. If the platform allows the feedback response, we also collect these responses; however, they were discarded in the analyses performed.

\begin{figure}[tb!] 
  \centering 
\includegraphics[width=.6\textwidth]{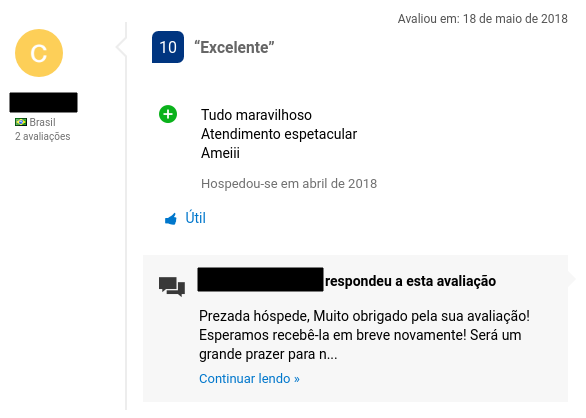} 
\caption{Illustration of a review on an online hosting service.} 
    \label{figComentarioAval} 
\end{figure}

Figure \ref{figComentarioAval} illustrates a review made on Booking. In this figure, the host responded to a review. Also, in this example, the word ``ameiii" (``loooved it" in English) does not exist formally in the Portuguese language, however people are free to use this type of construction, and this is not uncommon to be found in shared texts in social media. The tool for the sentiment analysis chosen understands and treats these cases. In the example, the word ``amei" was considered boosted, and this is reflected in the final sentiment evaluation of the review.

Between October 2016 and March 2017, we collected reviews from the platforms studied. Reviews can appear in several languages; however, we concentrate on written comments in English and Portuguese. After this filtering, we consider $648,030$ reviews from Booking, $115,760$ from Airbnb, and $8,589$ from CouchSurfing. Table \ref{tabDadosGeral} summarizes the data collected for each platform.

\begin{table}[tb] 
\centering 
\caption{Summary of dataset statistics for each platform considered.} 
\label{tabDadosGeral} 
\begin{tabular}{l|p{3cm}|p{3cm}} 
\hline 
             &  \# of accomodation & \# of reviews \\ \hline 
Booking      &       $880$         &         $648,030$           \\ 
Airbnb       &      $6,332$        &           $115,760$        \\ 
CouchSurfing &       $963$       &              $8,589$     
\end{tabular} 
\end{table}

%~~~~~~~~~~~~~~~~~~~~~~~~~~~~~~~~~~~~~~~~~~~~~~~~~~~~~~~~~~~~~~~~~~~~~~~~~~~~~~~~~~% 

\section{Sentiment Analysis}\label{secSent}

Sentiment analysis aims to extract opinions, sentiment, and emotions in different communication channels, mainly in the textual format \cite{narayanan2009sentiment,araujo2013measuring}, but also in other formats, such as in images \cite{CAMPOS201715,BONASOLI2020}. Particularly, the identification of sentiment in texts has become an important tool for the analysis of social media data, enabling several new services~\cite{liu2010sentiment,araujo2013measuring}. For example, companies can get users' opinions on the acceptance of a new product.

Current methods for detecting sentiment in sentences can be divided into two groups: based on machine learning, and based on lexical methods. Methods based on machine learning generally rely on a labeled database to train \cite{pang2002thumbs} classifiers, which can be considered a disadvantage due to the cost of obtaining labeled data. On the other hand, lexical methods use lists, dictionaries of words associated with specific sentiments. The efficiency of lexical methods is directly linked to the vocabulary used for the various contexts that exist. Hybrid approaches are also possible.

Several tools offer support to sentiment analysis, each one with particular characteristics. Abbasi et al. \cite{abbasi2014benchmarking} and Ribeiro et al. \cite {Ribeiro2016,araujo2016@icwsm} developed a benchmarking on several of these tools.  Authors demonstrated that the tool  SentiStrength \cite{thelwall2013heart} achieves high precision to analyze texts from social media, including reviews and other types of comments.

Sentistrengh uses a lexical dictionary labeled by humans that has been enhanced by machine learning. SentiStrength classifies the sentiment of content analyzed as positive or negative on a scale of $-4$ (strongly negative) to $+4$ (strongly positive), $0$ indicates neutral sentiment. This tool was chosen in this study because it presents better results in sentiment analysis for texts representing reviews in Web systems \cite{Ribeiro2016,abbasi2014benchmarking}.

Table \ref{tabExemploComen} illustrates examples of comments from our dataset and the polarity strength of the associated sentiment computed using the SentiStrength tool.

%%VINICIUS COMMENT: Esta tabela está tendenciosa, claramente os negativos são hoteis enquanto os positivos são dos airbnb ou couchsurfing
\begin{table*}[tb] 
\centering 
\scriptsize
\caption{Examples of comments and the associated strength of the sentiment polarity. Mentioned names have been replaced by X, Y and Z to preserve users' privacy.} 
\label{tabExemploComen} 
\begin{tabular}{p{1.7cm}|p{9.4cm}} 
\hline 
\textbf{Sentiment polarity} & \textbf{Review} \\ \hline 
-4                       &   Staff was extremely rude! ridiculously over priced, harsh and unwilling to assist, overall just not good.
 \\ \hline 
-3                       &   Very dirty and run down, tv remote coated in thick dust and the staff were so rude and unwelcoming.
         \\ \hline 
-2                       &  I am disappointed. They've checked in us into a dirty room, even the towels were not fresh and clean. So be careful and try to aware of staying there. 
        \\ \hline 
-1                       &  Run down hotel in desperate need of renewing
          \\ \hline 
0                        &  As shown in the pictures. Rooms could be bigger. \\ \hline 
1                        & Location was good. And the facilities were on point. \\ \hline 
2                        &   Stayed with X the first couple of days when I came to Rio. She is a great hostess, relaxed and very helpful. Recommended. \\ \hline 
3                        &  My second time in this apartment, now 5 days, again one of the best places I've rented, really great host Y and Z !!! 
          \\ \hline 
4                        &   Absolutely loved staying in Z's place in Copacabana!! Needless to say, the views are amazing. Great location - walking distance to many bars and restaurants. It's also close to the train. I would highly recommend it!       

\end{tabular} 
\end{table*}

%~~~~~~~~~~~~~~~~~~~~~~~~~~~~~~~~~~~~~~~~~~~~~~~~~~~~~~~~~~~~~~~~~~~~~~~~~~~~~~~~~~% 

\section{Sentiments in Reviews}\label{secResultados}

In this section, we present the analysis related to sentiment observed in the collected reviews.

\subsection{Aggregate Assessment}

\begin{figure}[tb!] 
  \centering 
\includegraphics[width=.6\textwidth]{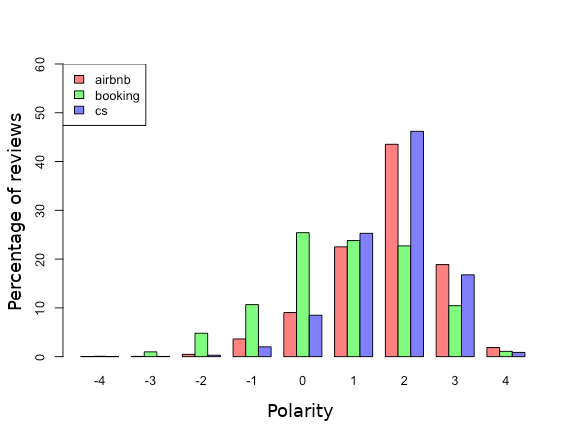} 
\caption{Distribution of sentiment for all platforms. In this figure the data are aggregated containing all cities without separation.} 
    \label{figSentGeral} 
\end{figure}

Figure \ref{figSentGeral} shows the distribution of sentiments for all platforms, Airbnb, Booking, and Couchsurfing (CS). The data considered are aggregated; that is, they contain all cities without separation. The $X$ axis represents the sentiment polarity for a particular review, and the $Y$ axis represents the percentage of reviews assigned to each polarity.

As we can see, platforms from sharing economy present more comments with positive polarity than in the platform from the traditional economy. The result presented in Figure \ref{figSentGeral} suggests that the type of economy can affect consumer sentiment in a review. Fradkin et al. \cite{fradkin2015bias} and Bulchand-Gidumal and Melián-González \cite{Gidumal2019} present some reasons that may explain the greater positivism or the lack of negativity in a review made on Airbnb. Among them: the personal interaction between host and guest often tends to occur; fear of receiving negative feedback from the host concerning the comment made; fear of harming the host with a poor rating that may discourage other potential guests.

\begin{figure}[tb!] 
\centering 
\subfigure[Curitiba] 
            {\includegraphics[width=.45\textwidth]{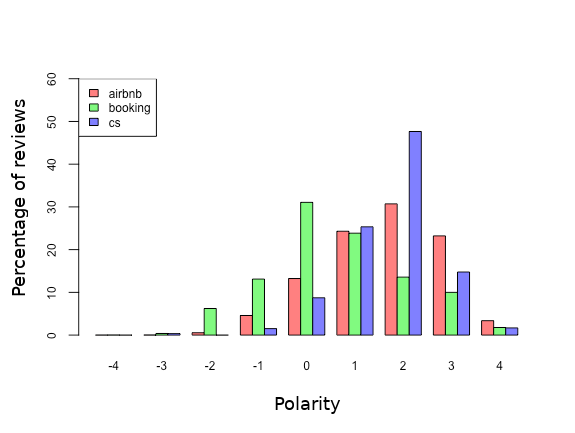}} 
\subfigure[Boston] 
            {\includegraphics[width=.45\textwidth]{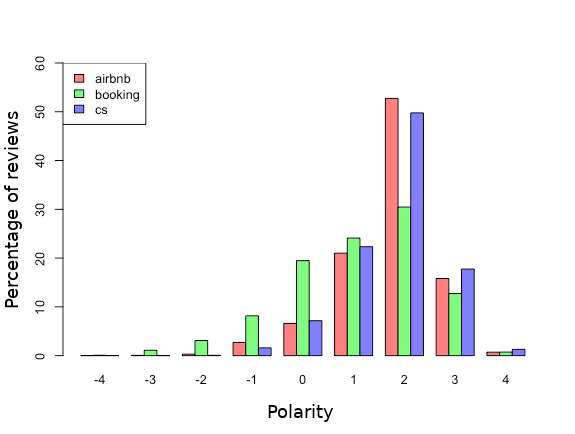}} 
\subfigure[Rio de Janeiro] 
            {\includegraphics[width=.45\textwidth]{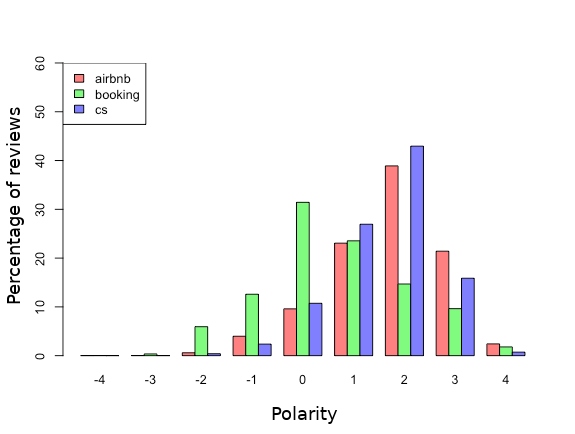}}
\subfigure[Las Vegas] 
            {\includegraphics[width=.45\textwidth]{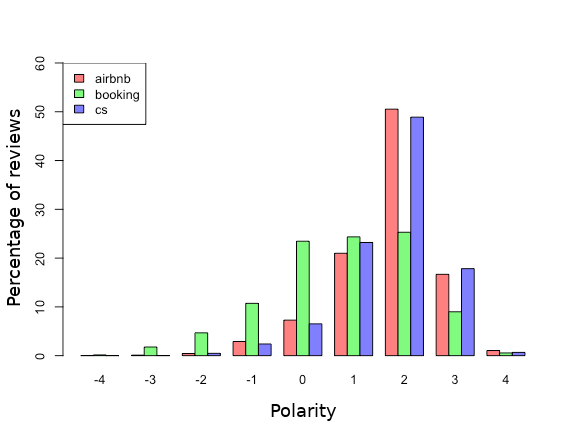}}
\subfigure[S\~ao Paulo] 
            {\includegraphics[width=.45\textwidth]{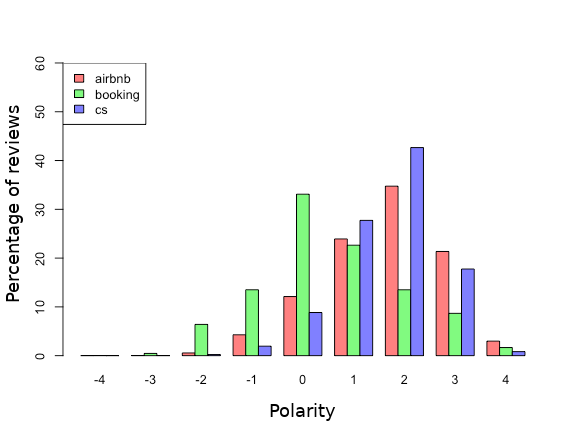}} 
\subfigure[New York] 
            {\includegraphics[width=.45\textwidth]{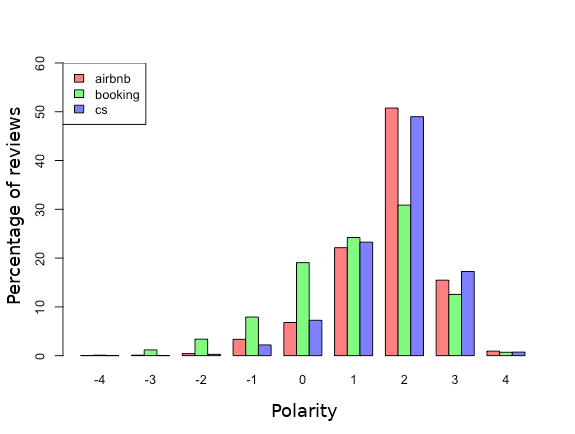}} 
\caption{Distribution of sentiment for all platforms, considering all cities studied separately.} 
\label{figCidadesBrasileiras} 
\end{figure}

Since Couchsurfing is a business with Airbnb-like relationships, these same motifs could also be used to explain the results found. Perhaps because it is free, some of these motives can still be leveraged, helping to explain the lower percentage of negative comments about Airbnb, mainly the reason for fear of receiving negative feedback from the host. In Couchsurfing, the user profile is very important to get accommodation. Users may be motivated not to perform inadequate evaluations of their hosts to avoid receiving any poor rating from them.

Also, in \cite{fradkin2015bias}, the authors note that in Airbnb, when a consumer is not satisfied with their experience, he/she tends not to write a comment, instead of negatively evaluating the place. This could also be explained by the above points.

At Booking, as the hosts are companies, typically hotels and often large properties, this sense of closeness between host and guest may not occur at the same frequency as in the other services analyzed. This might help explain the greater negativity in the comments when compared to Airbnb and Couchsurfing.

\subsection{Evaluation by Cities} \label{secAvaliacaoEntreCidades}

To check if there is any relationship between the city/country of the venue evaluated, we analyze each city considered separately. Figure \ref{figCidadesBrasileiras} shows the distribution of sentiments for all platforms, considering all cities studied separately. The $ X $ axis represents the sentiment polarity for a particular review, and the $Y$ axis represents the percentage of reviews assigned to each polarity.

With the help of this figure, it is possible to note that the country, or even the city alone, does not appear to be a relevant influence factor concerning the polarity of the reviews on the platforms. That is, a higher positivity regarding sharing economy accommodations is also seen when we separate the results by cities of the two countries analyzed. It is also possible to observe that the tendency for this disaggregated analysis is very similar to that found for the aggregate analysis, including the lower percentage of negative reviews seen in CouchSurfing compared to Airbnb.

Our results indicate that the presence of negative reviews is much smaller for online hosting services of sharing economy ($\approx$ 3\% of all comments: 4\% for Airbnb, and 2\% for Couchsurfing) compared to what is observed for the traditional economy ($\approx$ 17\% of all comments). This may hinder users' perception of the quality of a particular location. As negative evaluations tend to be more scarce in reviews in the sharing economy, neutral opinions can become more important. It is as if the scale of polarity began near the neutral, that is, representing the most negative opinions expressed by the users. This suggests that neutral evaluations should be taken into account at the time of choosing accommodation. These evaluations can perhaps make a difference in the classification and decision making when selecting a place to stay.

\subsection{Sentiments by Number of Comments}

One question that may arise at this point is whether the popularity of an establishment can influence the sentiment expressed by users on the analyzed platforms. To measure the intuition of popularity, we consider the number of comments that a given establishment received.

\begin{figure}[tb!] 
  \centering 
\includegraphics[width=.5\textwidth]{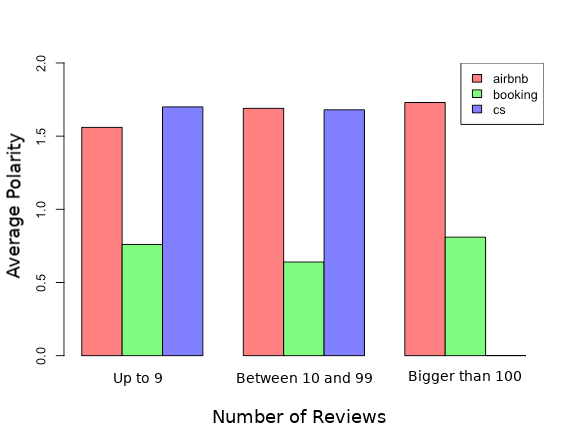} 
\caption{Relationship between the number of comments of an establishment by sentiment polarity in reviews.} 
    \label{figSentPopu} 
\end{figure}

Figure~\ref{figSentPopu} shows the relationship between the number of comments of a venue by sentiment polarity observed in reviews. The types of accommodations were grouped according to the number of reviews, up to $9$ reviews (unpopular), from $10$ to $99$ reviews (reasonably popular), over $100$ reviews (very popular). Please note that no CouchSurfing accommodation has received more than $100$ comments.

The result, presented by Figure \ref{figSentPopu}, suggests that the popularity of an establishment does not appear to have a significant influence on user opinion since the results for different venues groups according to their popularity did not vary considerably. This indicates that the results observed in Section \ref{secAvaliacaoEntreCidades} are not affected by venue popularity.

\subsection{Sentiment in Reviews on Homestays at Booking}  

In this section, we study whether the phenomenon observed (higher positivity in reviews of platforms of the sharing economy) is dependent on the platform. For that, we collected all the accommodations available in the category ``Home Stays" announced by Booking for all studied cities. This new collection was necessary because our previous dataset does not have the information on the type of accommodation.

Table \ref{tabHomeStayBooking} summarize the collected dataset. As expected, homestays at Booking are less available than hotel rooms. As we can see in the table, some accommodations did not have any review; several of them were recently added to the platform. We only considered accommodations with at least one review. For Las Vegas, this happened only one time; however, we have a reasonable number of reviews for this accommodation: $33$ reviews. In our dataset, S\~ao Paulo is the most popular city for this type of accommodation, having $27$ accommodations with reviews, presenting, in total, $670$ reviews. All data were collected between 5th - 10th January 2019.

\begin{table}[th]
\centering
\footnotesize
\caption{Summary of the dataset regarding ``Home Stays" on Booking for each studied city. }
\label{tabHomeStayBooking}
\begin{tabular}{lccc}
\hline
\textbf{City}  & \multicolumn{1}{l}{\textbf{Number of venues collected}} & \multicolumn{1}{l}{\textbf{With any review}} & \multicolumn{1}{l}{\textbf{Total of reviews}} \\ \hline
Boston         & 5                                                       & 3                                            & 51                                            \\
Curitiba       & 9                                                       & 4                                            & 105                                           \\
Las Vegas      & 3                                                       & 1                                            & 33                                            \\
New York       & 19                                                      & 7                                            & 53                                            \\
Rio de Janeiro & 64                                                      & 13                                           & 373                                           \\
S\~ao Paulo      & 62                                                      & 27                                           & 670                                          
\end{tabular}
\end{table}

\begin{figure}[tb!] 
\centering 
\subfigure[Curitiba] 
            {\includegraphics[width=.32\textwidth]{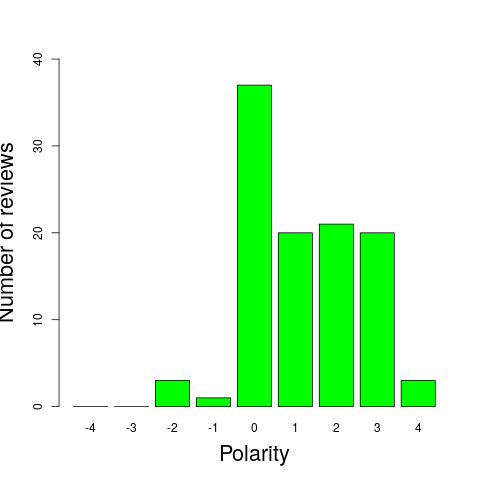}}  
\subfigure[Rio de Janeiro] 
            {\includegraphics[width=.32\textwidth]{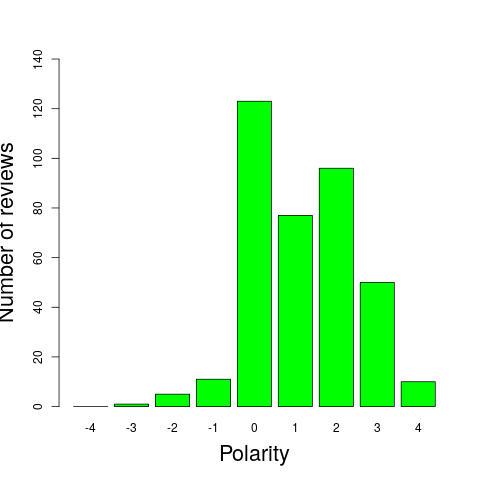}} 
\subfigure[S\~ao Paulo] 
            {\includegraphics[width=.32\textwidth]{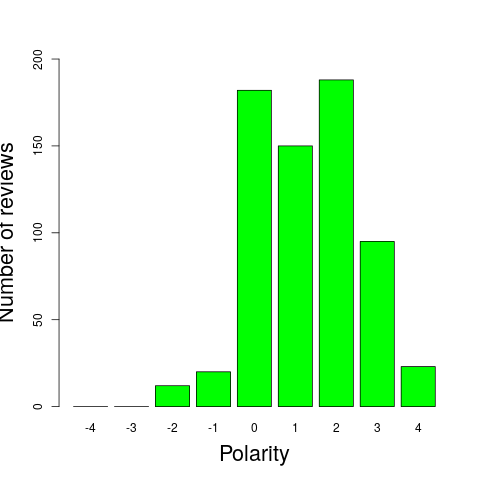}} 
\subfigure[Boston] 
            {\includegraphics[width=.32\textwidth]{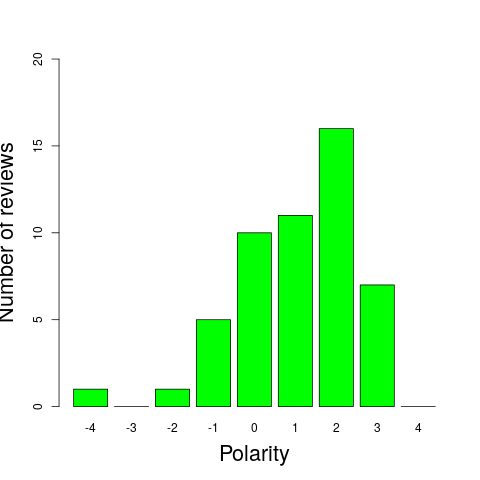}} 
\subfigure[Las Vegas] 
            {\includegraphics[width=.32\textwidth]{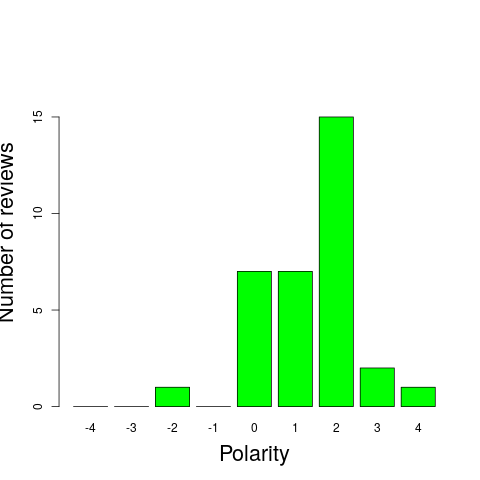}} 
\subfigure[New York] 
            {\includegraphics[width=.32\textwidth]{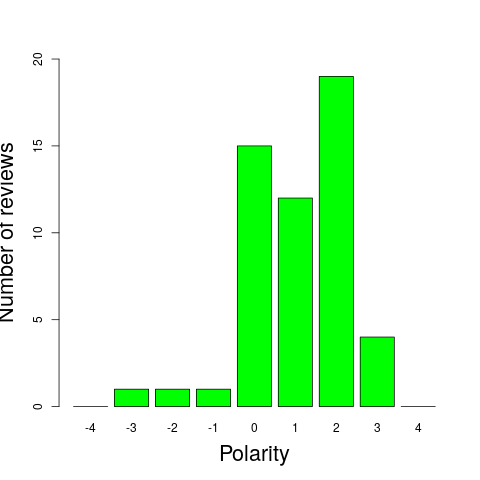}} 
\caption{Distribution of sentiment on reviews of homestays of Booking, considering all cities studied separately.}
\label{figSentHomestayBooking} 
\end{figure}

We applied the same methodology presented earlier to evaluate the sentiment of the reviews of this new dataset. Figure \ref{figSentHomestayBooking} shows the distribution of sentiment polarity for all cities. The $X$ axis represents the sentiment polarity for a particular review, and the $Y$ axis represents the number of reviews assigned to each polarity.

As we can observe with the help of this figure, the results are similar to the one presented in Figure \ref{figCidadesBrasileiras}, i.e., it was found very few negative reviews, and most of them are neutral or positive. We found that negative comments represent 4.9\% of the total, similarly to the proportion observed for Airbnb and Couchsurfing. This result also reinforces the hypothesis that the proximity between host and guest favors this sort of phenomenon.

\subsection{Content of Negative Comments}\label{secTopicNegativeEn}

In this section, we investigate the topics most addressed by users in negative comments. We focus on negative comments because we hypothesize that, in addition to rarer, they tend not to be very informative on hosting platforms of sharing economy. To do this analysis, we use the Latent Dirichlet Allocation (LDA) technique, a popular technique for modeling topics in textual content \cite{blei2003latent}.

Topical modeling is a method for unsupervised classification of documents similar. In our context, documents are reviews, containing different words. Particularly, LDA treats each document as a mixture of topics. This allows documents to ``overlap" each other in terms of content, rather than being separated into distinct groups, in order to mirror the typical use of natural language. For example, in a two-topic model, we could say that Review 1 is 85\% topic $X$ and 15\% topic $Y$, while Review 2 is 40\% topic $X$ and 60\% topic $Y$. In addition, LDA treats each topic as a mixture of words. Thus, consider a two-topic model for accommodation containing, for example, a topic for ``room" and another for ``breakfast". The most common words in the topic ``room" may be ``bed," ``sheet," and ``noise.". Whereas the topic ``breakfast" can be better represented by words like ``food", ``coffee" and ``juice". It is important to note that words can be shared between topics; a word like ``environment" may be expressive in both topics \cite{blei2003latent,silge2017text}.

Before identifying topics, we pre-process the reviews. We have removed URLs, special characters, unnecessary blank spaces, punctuations, numbers, stopwords, and performed a stemming process. After these steps, we identified $10$ topics for negative reviews (polarity $-4$ to $-1$) on Airbnb and Booking. Table \ref{tabTopicEN} presents ten words that best describe each of the topics. We note that all topics for Booking are negative. For example, Topic 2 is related to complaints regarding the room, and Topic 4 is more related to the staff. However, when analyzing the topics for Airbnb, we can identify several topics that suggest positive sentiments, all marked in bold and with ``**" in the table. For example, Topic 4 suggests being related to the accommodation in general, where the topic indicates that users have approved the stay.

\begin{table}[tb]
\centering 
\scriptsize
\label{tabTopicEN}
\caption{Ten latent topics from negative comments written in Engligh shared on Booking and Airbnb.} 
\begin{tabular}{llllllllll}
\hline
\multicolumn{10}{c}{\textbf{Topics for Booking}}                                                                                                                                                                                                                                    \\ \hline
\multicolumn{1}{c}{1} & \multicolumn{1}{c}{2} & \multicolumn{1}{c}{3} & \multicolumn{1}{c}{4}            & \multicolumn{1}{c}{5}            & \multicolumn{1}{c}{6}            & \multicolumn{1}{c}{7} & \multicolumn{1}{c}{8}            & \multicolumn{1}{c}{9} & \multicolumn{1}{c}{10} \\ \hline
hotel                 & room                  & check                 & staff                            & day                              & charg                            & breakfast             & like                             & room                  & bathroom               \\
stay                  & night                 & get                   & rude                             & call                             & book                             & poor                  & didnt                            & bed                   & dirti                  \\
park                  & door                  & time                  & desk                             & told                             & price                            & terribl               & just                             & small                 & clean                  \\
never                 & nois                  & elev                  & front                            & back                             & wifi                             & expens                & realli                           & smell                 & shower                 \\
will                  & sleep                 & wait                  & servic                           & ask                              & pay                              & coffe                 & even                             & smoke                 & old                    \\
locat                 & noisi                 & hour                  & one                              & said                             & use                              & pool                  & dont                             & uncomfort             & need                   \\
lot                   & air                   & peopl                 & help                             & got                              & hotel                            & servic                & look                             & view                  & water                  \\
better                & window                & long                  & guest                            & card                             & fee                              & food                  & can                              & chang                 & floor                  \\
money                 & work                  & took                  & recept                           & arriv                            & extra                            & area                  & one                              & bad                   & carpet                 \\
place                 & cold                  & one                   & custom                           & first                            & paid                             & bad                   & everyth                          & move                  & towel                  \\
                      &                       &                       &                                  &                                  &                                  &                       &                                  &                       &                        \\ \hline
\multicolumn{10}{c}{\textbf{Topics for Airbnb}}                                                                                                                                                                                                                                            \\ \hline
\multicolumn{1}{c}{1} & \multicolumn{1}{c}{2} & \multicolumn{1}{c}{3} & \multicolumn{1}{c}{\textbf{4**}} & \multicolumn{1}{c}{\textbf{5**}} & \multicolumn{1}{c}{\textbf{6**}} & \multicolumn{1}{c}{7} & \multicolumn{1}{c}{\textbf{8**}} & \multicolumn{1}{c}{9} & \multicolumn{1}{c}{10} \\ \hline
check                 & room                  & place                 & apart                            & hous                             & locat                            & host                  & everyth                          & work                  & clean                  \\
get                   & one                   & stay                  & good                             & like                             & walk                             & airbnb                & stay                             & didnt                 & bathroom               \\
time                  & bed                   & great                 & clean                            & make                             & close                            & book                  & home                             & day                   & kitchen                \\
arriv                 & night                 & worri                 & well                             & famili                           & easi                             & never                 & will                             & issu                  & shower                 \\
door                  & peopl                 & want                  & nice                             & feel                             & park                             & guest                 & need                             & use                   & dirti                  \\
even                  & bedroom               & littl                 & comfort                          & pictur                           & minut                            & hotel                 & help                             & just                  & floor                  \\
back                  & realli                & dont                  & new                              & much                             & area                             & experi                & recommend                        & problem               & towel                  \\
got                   & two                   & nice                  & build                            & made                             & street                           & first                 & time                             & water                 & provid                 \\
left                  & sleep                 & bit                   & lot                              & also                             & away                             & person                & perfect                          & night                 & use                    \\
host                  & live                  & look                  & locat                            & thing                            & quiet                            & list                  & friend                           & wifi                  & bed                   
\end{tabular}
\end{table}

The results for reviews in Portuguese follow a pattern similar to that presented for English. All topics for Booking are negative, and we can find several topics for Airbnb suggesting to be positive. These results are presented on \ref{appendixTopicPt}.
 
\section{Experiment with Volunteers}\label{secAvaliacaoUsuarios}

The analysis of reviews about venues before decision making is a task that is commonly performed by users of online hosting systems, such as those studied in this work \cite{6480220}. With that, based on the results of Section \ref{secResultados}, one question arises: based on reviews only, can users accurately rank accommodations of online hosting services of the sharing economy?

To evaluate this question, we have recruited $30$ volunteers with a diverse profile. In this group, we have representatives of various age groups, from adolescents to adults, having different levels of education, from incomplete higher education to postgraduate studies in progress.

Between February and April 2018, these volunteers were asked to respond to a questionnaire. In this questionnaire, volunteers needed to evaluate a particular venue only from the reviews made by other users. For this, we first chose four accommodations listed on Airbnb. These accommodations were selected with the aid of the mean sentiment polarity, as presented above. Taking into account the average polarity for each Airbnb establishment, the database of venues was divided into quartiles according to polarity. After this step, one venue was chosen randomly from each quartile. For each quartile, we assign the following labels, according to the polarity interval they represent: Q1 (quartile 1); Q2 (quartile 2); Q3 (quartile 3); and Q4 (quartile 4). Note that Q1 represents the best values and Q4 the worst.

We collected all available reviews for each selected venue. After a short text that serves to contextualize the volunteer, it should read the reviews of the venue and then classify it, assigning a note in a scale between 0 and 5, being 0 bad and 5 excellent. In the questionary, the order of presentation of the venues was scrambled. This was done to reduce some bias that the order of venues might bring in the respondents' perception. Table \ref{tabResultadosPesquisa} summarizes the $30$ responses provided by the volunteers.
  
\begin{table}[tb] 
\centering 
\footnotesize
\caption{Results of the experiment with users, containing the average score assigned by the volunteers for each venue.} 
\label{tabResultadosPesquisa} 
\begin{tabular}{cc}
\hline
\multicolumn{1}{c}{\textbf{Venue}} & \multicolumn{1}{c}{\textbf{Mean Score Given by the User (C.I. 95\%)}} \\ \hline 
Representative of Q1                  & $4.8$ ($\pm 0.15$)                               \\ 
Representative of Q2                        & $4$ ($\pm 0.31$)                                 \\ 
Representative of Q3                       & $4.6$ ($\pm 0.24$)                              \\ 
Representative of Q4                 & $3.4$ ($\pm 0.4$)       
\end{tabular} 
\end{table} 

The results of the experiment with users show that all venues were considered good or very good by the majority of users. For example, the venue representative of class Q4 had a good/average evaluation in the view of the volunteers, despite being in the quartile of venues with the lowest mean sentiment polarity.

This corroborates with the previously observed result, that is, in the sharing economy, which includes Airbnb, the reviews try to be more positive, which may hinder a human assessment. This suggests that venues with average polarity neutral (i.e., around zero polarity) may not be good accommodation options, it is as if the neutral is a negative polarity in this case. It is important to note that the representative venue of class Q3 has an average grade higher than the representative venue of class Q2. This experiment reinforces the suggestion that the perception of quality through reviews on Airbnb can be difficult.  

\section{Implications for the Design of New Functionalities}\label{secImplicacoes}

After analyzing the results presented in Sections \ref{secResultados} and \ref{secAvaliacaoUsuarios}, it is necessary to reflect on their possible implications. The greater positivity of Airbnb and Couchsurfing can be detrimental to consumers; after all, bad hosts may not be evident. Bad experiences that are not shared end up imposing difficulty in choosing accommodation. This fact can take the user to disregard a good place because of the ambiguity of sentiments present in the comments.

In some hosting services, such as Airbnb, users, in addition to doing reviews about their experience with the service obtained, can give a star rating, which can range from 1 (lowest) to 5 (highest ). However, as noted in an earlier study, 94\% of stars given by users on Airbnb range from $4.5$ to $5$ \cite{fradkin2015bias}. This makes this star rating not a sufficient metric for an analysis of the quality of a place to be done properly.

Our results suggest that it could be strategic to consider the polarity of reviews to help users in better decision making on hosting services of the sharing economy. Therefore, in this work, we exemplify a new way of evaluating these types of venues, taking into account the average polarity and the number of reviews of a given venue. For this, we suggest the following equation:

\begin{equation} 
score = \log{C} + (4+P)^2, 
\end{equation}

where $P \in [-4, 4]$ represents the mean polarity of the venue's reviews, and $C \geq 0$ represents the number of reviews for a venue.

Thus, a higher $score$ value tends to be attributed to venues with a more positive average sentiment in the reviews. The equation also takes into account the number of evaluations performed, where the more opinions, the better. By taking into account the number of reviews, it is possible to distinguish whether venues have enough reviews for an evaluation.

For each Airbnb venue collected, a $score$ was assigned considering the equation above. \ref{appendixScore} shows some examples of $score$ for this analysis, helping to understand how each variable impacts $score$. For the venues studied in Section \ref{secAvaliacaoUsuarios} (shown in the Table \ref{tabResultadosPesquisa}), $ score $ would rank the locations as follows: 1st Q1 ($ score $ = 51.07); 2nd Q2 ($score$ = 40.66); 3rd Q3 ($score$ = 34.91); 4th Q4 ($score$ = 14.04).

We believe $score$, or some variation in this direction, can be useful, for example, to design a new venue ranking. Just as there are already rankings from the lowest to the highest price, a ranking could be created according to the calculated $score$. This may perhaps help users make better decisions when choosing a place to stay. A qualitative assessment with users to see if this approach can help improve the user experience is outside the scope of this work; however, it is important to be carried out.

%~~~~~~~~~~~~~~~~~~~~~~~~~~~~~~~~~~~~~~~~~~~~~~~~~~~~~~~~~~~~~~~~~~~~~~~~~~~~~~~~~~% 

\section{Conclusion and Future Work}\label{secConclusao}

By evaluating reviews on two hosting platforms of the sharing economy and one of the traditional economy, we find evidence that reviews tend to be more positive on platforms of the sharing economy. This corroborates with the hypothesis that this phenomenon happens due to personal contact that occurs between the host and the guest on those services. This result can be detrimental to consumers as bad hosts may not be evident. More importantly, our findings suggest that reviews on different platforms might require different interpretations, especially for algorithmic decision-making approaches that use review datasets in the learning phase. We hope our quantitative analysis and observations may inspire new approaches able to account for this perceived bias towards positivity in hosting platforms.

As future work, we intend to study patterns between sentiment in reviews with other attributes, such as gender and time. In addition, we want to conduct a qualitative assessment to investigate whether new ways to rank venues exploring the insights obtained in this study, such as the approach presented, can bring a better experience for users.

 %==================================================================================================
 
 \begin{acknowledgements}

This study was financed in part by the Coordenação de Aperfeiçoamento de Pessoal de Nível Superior - Brasil (CAPES) - Finance Code 001. This work is also partially supported by the project URBCOMP (Grant \#403260 /2016-7 from National Council for Scientific and Technological Development agency - CNPq) and GoodWeb (Grant \#2018/23011-1 from Sao Paulo Research Foundation - FAPESP). The authors would also like to thank Marcelo Santos and all the volunteers for the valuable help in this study.

\end{acknowledgements}

 \section*{Conflict of interest}

 The authors declare that they have no conflict of interest.

\bibliographystyle{spmpsci}      % mathematics and physical sciences

%\begin{thebibliography}{}
% etc
%\end{thebibliography}

 \bibliography{references}

\clearpage

\appendix

\section{Topics of Negative Comments in Portuguese}\label{appendixTopicPt}

This section presents the topic analysis for negative reviews written in Portuguese, following the same methodology presented in Section \ref{secTopicNegativeEn}. The results in Table \ref{tabTopicPt} follow a similar pattern to the one observed English reviews. 

\begin{table}[h]
\centering 
\tiny
\caption{Ten latent topics from negative comments written in Engligh shared on Booking and Airbnb.} 
\begin{tabular}{llllllllll}
\label{tabTopicPt}
\\ \hline
\multicolumn{10}{c}{\textbf{Topics for Booking}}\\ 
\hline
\multicolumn{1}{c}{1} & \multicolumn{1}{c}{2} & \multicolumn{1}{c}{3} & \multicolumn{1}{c}{4} & \multicolumn{1}{c}{5} & \multicolumn{1}{c}{6} & \multicolumn{1}{c}{7}            & \multicolumn{1}{c}{8} & \multicolumn{1}{c}{9}            & \multicolumn{1}{c}{10}            \\ \hline
banheir               & estacion              & barulh                & pouc                  & falt                  & cam                   & quart                            & ruim                  & nao                              & condic.                         \\
pequen                & hotel                 & tod                   & ser                   & quart                 & quart                 & mal                              & wifi                  & dia                              & hotel                             \\
chuveir               & car                   & rua                   & pod                   & limpez                & suj                   & cheir                            & atend                 & reserv                           & velh                              \\
port                  & demor                 & noit                  & caf                   & deix                  & toalh                 & mof                              & nad                   & ped                              & elev                              \\
box                   & check                 & fic                   & manh                  & tom                   & roup                  & corredor                         & internet              & vez                              & antig                             \\
quebr                 & pag                   & outr                  & piscin                & sujeir                & trave.            & carpet                           & frac                  & hosped                           & funcion                           \\
defeit                & cobr                  & andar                 & ter                   & frigob                & casal                 & recepcion                        & tud                   & cheg                             & barulh.                         \\
vazament              & pouc                  & janel                 & restaur               & problem               & troc                  & equip                            & sinal                 & pra                              & precis                            \\
agu                   & pesso                 & faz                   & melhor                & desej                 & apen                  & esgot                            & lent                  & hav                              & reform                            \\
pia                   & hor                   & apart                 & dev                   & muit                  & solteir               & cigarr                           & lig                   & volt                             & parec                             \\
                      &                       &                       &                       &                       &                       &                                  &                       &                                  &                                   \\ \hline
\multicolumn{10}{c}{\textbf{Topics for Airbnb}}                                                                                                                                                                                                                                 \\ \hline
\multicolumn{1}{c}{1} & \multicolumn{1}{c}{2} & \multicolumn{1}{c}{3} & \multicolumn{1}{c}{4} & \multicolumn{1}{c}{5} & \multicolumn{1}{c}{6} & \multicolumn{1}{c}{\textbf{7**}} & \multicolumn{1}{c}{8} & \multicolumn{1}{c}{\textbf{9**}} & \multicolumn{1}{c}{\textbf{10**}} \\ \hline
recom                 & brev                  & recom                 & apart.             & estad                 & excelent              & pen                              & tud                   & volt                             & nov                               \\
val                   & volt                  & aconcheg              & estad                 & apart.             & tud                   & tod                              & lug                   & fic                              & hosped                            \\
pod                   & obrig                 & real                  & suj                   & cert                  & bem                   & recom                            & apart.             & apart.                        & simpl                             \\
med                   & certez                & alug                  & limp                  & tud                   & gost                  & certez                           & banheir               & pont                             & dia                               \\
bem                   & cas                   & nao                   & recom                 & otim                  & pert                  & localiz                          & atenc                 & sempr                            & otim                              \\
pra                   & rio                   & cam                   & quart                 & limpez                & banheir               & atend                            & problem               & pesso                            & trab                              \\
esper                 & val                   & tir                   & funcion               & esper                 & deix                  & ador                             & por                   & esper                            & volt                              \\
realment              & best                  & ache                  & fot                   & nao                   & problem               & curt                             & receb                 & pert                             & pod                               \\
sent                  & barulh                & bem                   & dias                  & cama                  & entend                & val                              & qualqu                & obrig                            & espac                             \\
ador                  & quart                 & resolv                & forte.             & precis                & piscin                & bom                              & sempr                 & recom                            & bom                              
\end{tabular}
\end{table}

\section{Simulation of Score}\label{appendixScore}

Table \ref{tabExemplosNota} shows some examples of $score$ based on reviews and polarity of Airbnb. This help us to have an idea on how each variable impacted $ score$. In these examples, the $ score $ values ranged from $ 5.09$ (worst $score$) to $64.69$ (best $score$).

\begin{table}[htb] 
\centering 
\footnotesize
\caption{Examples of $ score $ based on reviews and polarity of Airbnb.} 
\label{tabExemplosNota} 
\begin{tabular}{ccc} 
\cline{1-3} 
\# Reviews & Mean polarity of sentiments & $score$     \\ \hline 
2              & 4                                & 64.69 \\ 
8              & 3                                & 51.08 \\ 
3              & 3                                & 50.10 \\ 
19             & 2.8333                           & 49.64 \\ 
40             & 2.0938                           & 40.82 \\ 
26             & 1.913                            & 38.22 \\ 
2              & 1.5                              & 30.94 \\ 
11             & 0.2                              & 20.04  \\ 
13               & -1                                  & 11.56 \\
3              & -1                               & 10.10 \\
3               & -2                                 &    5.09    
\end{tabular} 
\end{table}

\end{document}